\documentclass[11pt,a4paper]{article}

\usepackage[utf8]{inputenc}
\usepackage[T1]{fontenc}
\usepackage{lmodern}
\usepackage[margin=25mm]{geometry}
\usepackage{amsmath,amssymb}
\usepackage{graphicx}
\usepackage{booktabs}
\usepackage{multirow}
\usepackage{siunitx}
\usepackage{xcolor}
\usepackage[colorlinks=true,linkcolor=blue!60!black,citecolor=blue!60!black,urlcolor=blue!60!black]{hyperref}
\usepackage[numbers,sort&compress]{natbib}
\usepackage{authblk}
\usepackage[section]{placeins}

\graphicspath{{figures/}}

\DeclareSIUnit{\angstrom}{\text{\AA}}
\newcommand{\rtwoscan}{r$^2$SCAN}

\title{Technical Note on Matlantis PFP v9:\\ Extending the Element and Domain Coverage\\ of the \texorpdfstring{r$^2$SCAN}{r2SCAN} Training Data}

\author[1]{Daiki Shintani}
\author[1]{Shigeru Iwase}
\author[1]{Jun Oshiki}
\author[1]{Kohei Shinohara}
\author[1]{Nontawat Charoenphakdee}
\author[1]{Makoto Hiramatsu}
\author[1]{Yuta Tsuboi}
\author[2]{Yusuke Asano}
\author[1]{Hirotaka Akita}
\author[1]{Yong-Bin Zhuang}
\author[1]{So Takamoto}
\author[1]{Chikashi Shinagawa}
\affil[1]{Preferred Networks, Inc., Tokyo, Japan}
\affil[2]{Matlantis Corporation, Tokyo, Japan}

\date{November 18, 2026}

\begin{document}
\maketitle

\begin{abstract}
Universal machine learning interatomic potentials (uMLIPs) make atomistic simulations possible at scales beyond the reach of density functional theory (DFT).
Matlantis PFP is a uMLIP that treats a variety of domains, including crystals, molecules, and surfaces, with a single model.
PFP v8 incorporated a training dataset computed with the \rtwoscan{} functional, chosen for its good agreement with experiment, but the elements and domains covered by that dataset were limited.
Here we present Matlantis PFP v9, in which the training data computed with the \rtwoscan{} functional are extended to the lanthanides from Pr to Lu and to the elements from Po to Cm, for a total of 96 elements, and to additional domains such as clusters and adsorption structures.
The formation energies of crystals containing the newly supported elements reproduce experimental standard enthalpies of formation with the same level of accuracy as crystals composed only of previously supported elements, and the adsorption energies of molecules on transition-metal surfaces have an overall error against experiment comparable to that of DFT-PBE, although chemisorbed species are overbound.
On the same type of GPU (A100), the inference speed and the maximum system size exceed those of SevenNet-Omni, a public uMLIP trained on \rtwoscan{} data: forces are evaluated three to four times faster, and five to ten times as many atoms can be handled.
As a result, PFP-R2SCAN is now applicable to systems containing rare-earth and actinide elements and to adsorption on surfaces.
\end{abstract}

\section{Introduction}
\label{sec:introduction}

Machine learning interatomic potentials (MLIPs) learn from reference data computed with density functional theory (DFT) and provide interatomic interactions of near-DFT accuracy at a far lower computational cost.
They have made it possible to run molecular dynamics on larger systems and over longer timescales than DFT allows, up to 100 million atoms with ab initio accuracy~\cite{jia2020deepmd}.
They have also made it possible to search exhaustively over large numbers of candidate structures, which has led to the discovery of millions of new stable crystals~\cite{merchant2023gnome}.
Early MLIPs were restricted to specific targets, such as organic molecules built from a handful of elements~\cite{smith2017ani}.
Universal MLIPs (uMLIPs), which treat most of the periodic table and a wide variety of materials with a single model, have since been developed, and several are now publicly available~\cite{chen2022m3gnet,deng2023chgnet,batatia2025mace,wood2025uma,kim2026sevennetomni}.

Matlantis PFP (hereafter PFP) is a uMLIP designed to treat a variety of domains, including crystals, molecules, surfaces, and adsorption structures, with a single model rather than a separate model for each domain~\cite{takamoto2022pfp}.
PFP learns several datasets computed with different functionals simultaneously, using a calculation-mode label, and has from its first version combined crystal data computed with the PBE functional~\cite{perdew1996pbe} and molecular data computed with the $\omega$B97X-D functional~\cite{chai2008wb97xd}.
In PFP v8, a dataset computed with the \rtwoscan{} functional~\cite{furness2020r2scan} was added with agreement with experiment as an explicit design target, and PFP v8 outperformed PBE-based models in reproducing experimental formation energies of crystals, reaction energies of molecules, surface energies, and melting points~\cite{shinagawa2026pfpv8}.
However, this dataset covered only 70 elements and fewer domains in PFP v8 than the dataset computed with the PBE functional.
Most of the lanthanides were left out because the choice of PAW potentials for the $f$-block elements was being re-examined with the move to the \rtwoscan{} functional.
It had not yet been settled whether to carry over the potentials of the dataset computed with the PBE functional, which for some elements keep the 4$f$ electrons in the frozen core, or to treat the 4$f$ electrons as valence electrons.
The elements from Po to Cm were left out because of their lower priority for the intended applications, and the domains because \rtwoscan{} calculations cost about four times as much as PBE calculations.

This technical note on PFP v9 reports how the element and domain coverage of the training data computed with the \rtwoscan{} functional was extended, together with the results of its validation.
Section~\ref{sec:method} describes the training datasets and the DFT conditions for the newly supported elements, Section~\ref{sec:results} presents the validation, and Section~\ref{sec:conclusion} concludes with directions for future work.
\section{Method}
\label{sec:method}

The changes from PFP v8~\cite{shinagawa2026pfpv8} to PFP v9 are confined to the training data.
The model architecture, the training procedure, and the inference engine are unchanged, and we refer the reader to the PFP v8 paper for these.
Below we describe the training datasets and the DFT conditions used for the newly supported elements.

\subsection{Training datasets}
\label{sec:datasets}

The training data of PFP consist of several datasets computed with different functionals.
The dataset computed with the PBE functional, part of which includes a Hubbard U correction~\cite{dudarev1998u} (hereafter the PBE/+U dataset), and the molecular dataset computed with the $\omega$B97X-D functional are unchanged from PFP v8.
The dataset computed with the \rtwoscan{} functional was introduced in PFP v8 and is extended in PFP v9.
Table~\ref{tab:datasets} lists the domains, the number of elements, and the number of data points in each dataset.
At inference, the user selects a calculation mode that specifies which of these datasets the model reproduces: PFP-R2SCAN for the \rtwoscan{} data, PFP-PBE and PFP-PBE+U for the PBE/+U data, and PFP-WB97XD for the $\omega$B97X-D data.
The PBE-based and \rtwoscan{}-based modes are also available with the DFT-D3 dispersion correction with Becke--Johnson damping~\cite{grimme2010d3,grimme2011bj}, denoted by the suffix +D3; PFP-WB97XD has no such variant because the $\omega$B97X-D functional already includes a dispersion correction.

\begin{table}[htbp]
  \centering
  \caption{Datasets used to train PFP v9. Each dataset is named after the functional used to compute it. Entries in bold differ from PFP v8: ``new'' marks a domain added in PFP v9, and the numbers of elements and data points were 70 and $3\times10^{6}$ in PFP v8.}
  \label{tab:datasets}
  \small
  \setlength{\tabcolsep}{4pt}
  \begin{tabular}{lccccccrr}
    \toprule
    & \multicolumn{6}{c}{Domains} & & \\
    \cmidrule(lr){2-7}
    Dataset & Molecule & Bulk & Slab & Disorder & Cluster & Adsorption & Elements & Data \\
    \midrule
    \rtwoscan{} & \checkmark & \checkmark & \checkmark & \checkmark & \textbf{new} & \textbf{new} & \textbf{96} & {\boldmath$6\times10^{6}$} \\
    PBE/+U & \checkmark & \checkmark & \checkmark & \checkmark & \checkmark & \checkmark & 96 & $6\times10^{7}$ \\
    $\omega$B97X-D & \checkmark & & & & & & 9 & $6\times10^{6}$ \\
    \bottomrule
  \end{tabular}
\end{table}

In PFP v8, the dataset computed with the \rtwoscan{} functional covered four domains, namely molecules, bulk, slabs, and disordered structures, for 70 elements.
The disordered structures are generated by running high-temperature molecular dynamics on randomly chosen combinations of elements; they lie far from real materials but are included to improve generalization across diverse chemical environments.
PFP v9 extends the dataset in two directions.
The first is the elements: the 13 lanthanides from Pr to Lu and the 13 elements from Po to Cm were added, bringing the coverage to the same 96 elements as the PBE/+U dataset.
The second is the domains: a variety of domains missing from the dataset in PFP v8, including clusters and adsorption structures, were added, so that the dataset now covers the same domains as the PBE/+U dataset.
The number of data points increased from $3\times10^{6}$ in PFP v8 to $6\times10^{6}$.
In addition to these datasets, the OC20 dataset~\cite{chanussot2021oc20}, a public dataset of adsorption structures computed with the RPBE functional, is used as auxiliary data during training, as in PFP v8; it is not offered as a calculation mode at inference and is omitted from Table~\ref{tab:datasets}.

\subsection{DFT calculation conditions}
\label{sec:pseudopotentials}

The data newly added in PFP v9 were computed under the same conditions as the existing data computed with the \rtwoscan{} functional, including the plane-wave cutoff energy, the k-point density, the SCF convergence criterion, and the treatment of magnetic states.
These conditions are described in the PFP v8 paper~\cite{shinagawa2026pfpv8}.
For the newly supported elements, we selected the pseudopotentials as follows.

\begin{figure}[htbp]
  \centering
  \includegraphics[width=\linewidth]{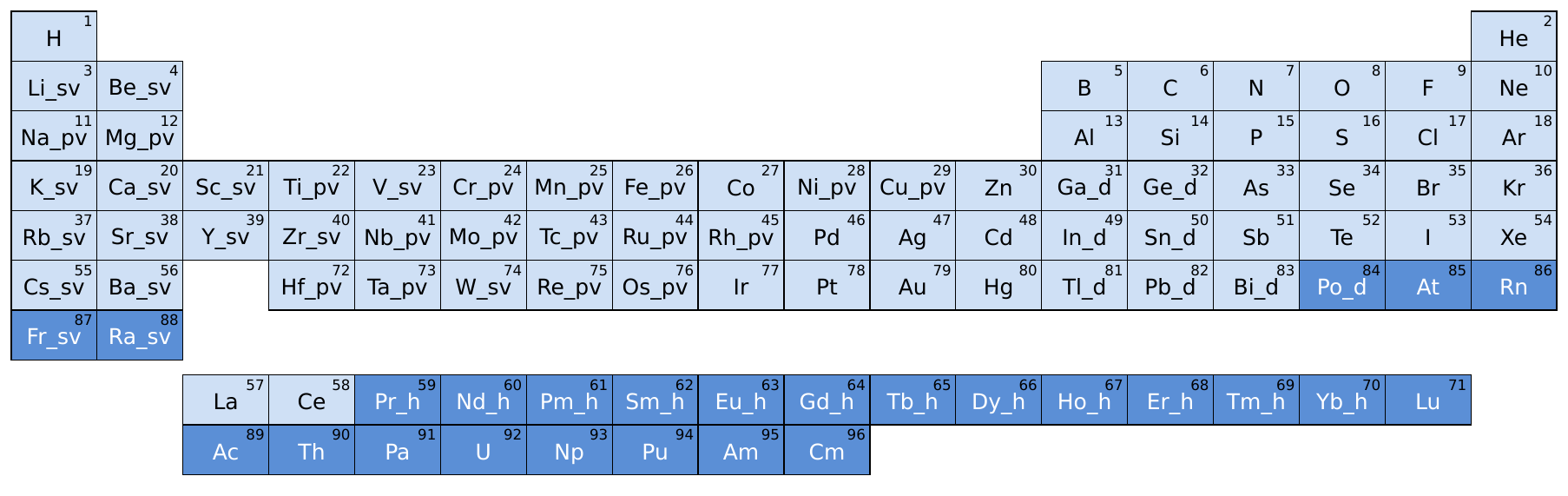}
  \caption{PAW pseudopotentials (\texttt{potpaw\_PBE.64}) used in the calculations with the \rtwoscan{} functional. The labels are the VASP potential names; a label without a suffix denotes the standard potential. Dark cells mark the elements newly supported in PFP v9 and light cells those supported since PFP v8.}
  \label{fig:pseudopotentials}
\end{figure}

As in PFP v8, we used the latest standard set of PAW potentials~\cite{kresse1999paw} for PBE distributed with VASP~\cite{kresse1996vasp} (\texttt{potpaw\_PBE.64}).
For the lanthanides, two kinds of PAW potentials are available in this set: those that keep the 4$f$ electrons in the frozen core and thereby fix the formal valence (for example, the suffix \texttt{\_3} for the trivalent state), and hard potentials (with the suffix \texttt{\_h}) that treat all 4$f$ electrons as valence electrons.
The frozen-core potentials are numerically stable and have been used in most PBE-based databases, including the Materials Project~\cite{materialsproject}, whose choices the PBE dataset of PFP follows~\cite{takamoto2022pfp}, but they cannot describe systems in which the valence changes or in which the 4$f$ orbitals take part in bonding.
The hard potentials were prepared for a cross-code verification project of DFT precision~\cite{bosoni2024acwf}, validated against all-electron calculations, and are adopted by recent \rtwoscan{} training datasets such as MatPES~\cite{kaplan2025matpes} and MP-ALOE~\cite{mpaloe2025}.
PFP v9 uses the hard potentials for the lanthanides from Pr to Yb, so that a single potential applies regardless of the oxidation state.
For Lu, At, Rn, and the actinides from Ac to Cm we used the standard potentials; for Po, the potential that includes the $d$ electrons in the valence; and for Fr and Ra, the potentials that include the semicore $s$ electrons in the valence.
The choices for the 70 previously supported elements are unchanged from PFP v8.
Figure~\ref{fig:pseudopotentials} shows the pseudopotentials used for all 96 elements in the calculations with the \rtwoscan{} functional.

\section{Results}
\label{sec:results}

\subsection{Formation energies of crystals containing the newly supported elements}
\label{sec:new-elements}

For crystals containing the newly supported elements, we compared the formation energies computed with PFP-R2SCAN against experimental standard enthalpies of formation.
The experimental values were taken from the experimental formation enthalpy dataset compiled by Wang et al.~\cite{wang2021dftcorrections} and distributed with Matminer~\cite{matminer}.
Of the 26 elements newly supported in PFP v9, 15 appear in this dataset: Pr, Nd, Sm, Eu, Gd, Tb, Dy, Ho, Er, Tm, Yb, Lu, Th, U, and Pu.
For each system, the corresponding structure from the Materials Project~\cite{materialsproject} was used as the initial structure, both the cell and the atomic positions were relaxed with PFP-R2SCAN, and the formation energy was evaluated.
The relaxation used the FIRE optimizer~\cite{bitzek2006fire} with the Fr\'echet cell filter of ASE~\cite{larsen2017ase}, a convergence criterion of \SI{0.005}{eV/\angstrom} on the maximum force, and at most 1,000 steps.
The elemental reference energies were obtained by relaxing, under the same conditions, the most stable structure of each element from the \rtwoscan{} calculations registered in the Materials Project~\cite{kingsbury2022}.
Temperature, pressure, and zero-point vibration are not taken into account, so the comparison is between 0~K formation energies and standard enthalpies of formation.

\begin{figure}[tbp]
  \centering
  \includegraphics[width=\linewidth]{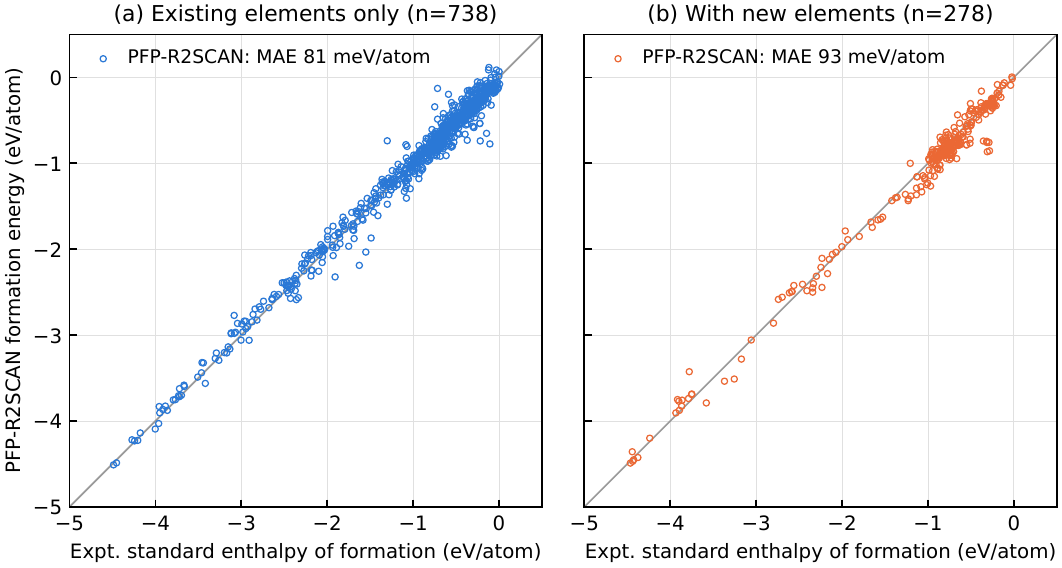}
  \caption{PFP-R2SCAN formation energies of crystals against experimental standard enthalpies of formation, for (a) crystals composed only of previously supported elements and (b) crystals containing at least one element newly supported in PFP v9. The legend gives the MAE against experiment, and $n$ is the number of systems.}
  \label{fig:formation-energy}
\end{figure}
Figure~\ref{fig:formation-energy} compares the PFP-R2SCAN formation energies with experiment, separately for crystals composed only of previously supported elements and for crystals containing newly supported elements.
The MAE is \SI{81}{meV/atom} for the former and \SI{93}{meV/atom} for the latter; the latter is slightly larger, but the two are at the same level.
For reference, the DFT \rtwoscan{} formation energies registered in the Materials Project give MAEs of \SI{80}{meV/atom} and \SI{84}{meV/atom}, respectively, against the same experimental values for the same systems.
\FloatBarrier

\subsection{Adsorption energies of molecules on transition-metal surfaces}
\label{sec:adsorption}

As a validation corresponding to the added adsorption domain, we compared the adsorption energies of molecules on transition-metal surfaces against experiment.
The benchmark is the database of adsorption energies compiled by Wellendorff et al.~\cite{wellendorff2015adsorption}.
It provides experimental adsorption energies measured on single-crystal surfaces with the zero-point vibrational contribution removed, so that they can be compared directly with static calculations.
The target comprises 36 systems: 23 classified as chemisorption and 13 dominated by dispersion interactions.
The former include molecular adsorption of CO and NO on Ni, Pt, Pd, Rh, Ir, Cu, Ru, and Co surfaces, atomic adsorption of O, H, and I, and the dissociative adsorption of NO into N and O on Ni(100).
The latter include NH$_3$ on Cu(100) and methane, ethane, propane, butane, benzene, cyclohexene, methanol, water, and methyl iodide on noble-metal surfaces, mainly Pt(111).
The dissociative adsorption was modeled as co-adsorption of an N atom and an O atom on separate neighboring sites and compared per NO molecule.
Of the 39 surface reactions in the database, the following three were excluded.
The dissociation of CH$_2$I$_2$ into CH, H, and two I on Pt(111) and the dissociation of CH$_3$I into CH$_3$ and I on Pt(111) produce several adsorbed species including polyatomic fragments, and the experiments were performed at low coverages of 1/12 and 1/25 per surface atom.
Reproducing these coverages requires surface cells much larger than the $2\times2$ and $3\times3$ cells used for the other systems.
The combinations of relative positions that the fragments can take in such cells cannot be exhausted by the site-scanning procedure described below, so these two systems were excluded.
The adsorption of H$_2$O on Pt(111) pre-covered with O atoms, which forms a hydrogen-bonded network of H$_2$O and OH, is a reaction with a pre-adsorbed species rather than adsorption on a clean surface and cannot be treated by the same procedure.

The clean surface, the isolated adsorbate, and the adsorption structure were each optimized with PFP-R2SCAN to obtain the adsorption energy. To see the effect of the DFT-D3 dispersion correction, the same procedure was repeated with PFP-R2SCAN+D3.
The optimization was converged to a maximum force of \SI{0.001}{eV/\angstrom}.
For the adsorption site, optimizations were started from all candidate high-symmetry sites of the facet and the most stable structure was adopted.
For the dissociative adsorption, optimizations were started from every combination of candidate sites for the N atom and the O atom.
For atomic adsorption referenced to O$_2$, H$_2$, and I$_2$, the energies were converted to values per adsorbed atom.

\begin{figure}[tb]
  \centering
  \includegraphics[width=\linewidth]{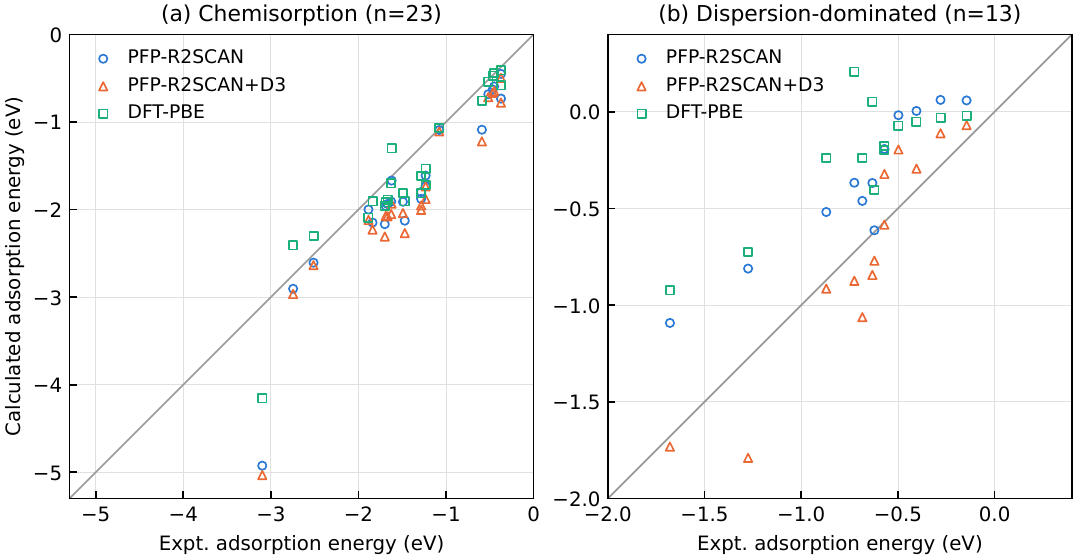}
  \caption{Adsorption energies of molecules on transition-metal surfaces against experiment, for (a) the 23 chemisorption systems and (b) the 13 dispersion-dominated systems. The DFT-PBE values are those reported by Wellendorff et al.~\cite{wellendorff2015adsorption}, and $n$ is the number of systems.}
  \label{fig:adsorption-energy}
\end{figure}
\begin{table}[tb]
  \centering
  \caption{Errors of the adsorption energies of molecules on transition-metal surfaces against experiment, for 23 chemisorption systems and 13 dispersion-dominated systems. The DFT-PBE values compare the PBE calculations reported by Wellendorff et al.~\cite{wellendorff2015adsorption} against the same experimental values.}
  \label{tab:adsorption}
  \small
  \begin{tabular}{lrrrrrr}
    \toprule
    & \multicolumn{2}{c}{All (meV)} & \multicolumn{2}{c}{Chemisorption (meV)} & \multicolumn{2}{c}{Dispersion-dominated (meV)} \\
    \cmidrule(lr){2-3}\cmidrule(lr){4-5}\cmidrule(lr){6-7}
    Mode / functional & MAE & RMSE & MAE & RMSE & MAE & RMSE \\
    \midrule
    PFP-R2SCAN & 354 & 464 & 361 & 510 & 342 & 369 \\
    PFP-R2SCAN+D3 & 364 & 498 & 465 & 598 & 186 & 232 \\
    DFT-PBE & 333 & 416 & 254 & 341 & 472 & 521 \\
    \bottomrule
  \end{tabular}
\end{table}
Figure~\ref{fig:adsorption-energy} compares the PFP-R2SCAN and PFP-R2SCAN+D3 adsorption energies with experiment, and Table~\ref{tab:adsorption} gives their errors together with those of the DFT-PBE calculations reported in the same database.
Over all 36 systems, the MAE is \SI{354}{meV} for PFP-R2SCAN and \SI{364}{meV} for PFP-R2SCAN+D3, comparable to the \SI{333}{meV} of DFT-PBE.
The two groups of systems behave differently, however.
For the dispersion-dominated systems, PFP-R2SCAN underbinds all 13 systems, with a mean error of \SI[retain-explicit-plus]{+342}{meV}, as does DFT-PBE with \SI[retain-explicit-plus]{+472}{meV}; the D3 correction adds 0.13 to 0.98~eV of binding and lowers the MAE to \SI{186}{meV}.
For the chemisorption systems, PFP-R2SCAN overbinds all 23 systems, with a mean error of \SI{-361}{meV} against \SI{-176}{meV} for DFT-PBE; the D3 correction adds a further 0.02 to 0.26~eV of binding and raises the MAE to \SI{465}{meV}, above the \SI{254}{meV} of DFT-PBE.
The largest error, \SI{1.8}{eV}, is for the dissociative adsorption of NO on Ni(100), which DFT-PBE also overbinds by \SI{1.1}{eV}.
\FloatBarrier

\subsection{Inference speed and maximum number of atoms}
\label{sec:speed}

We compared the inference speed and the maximum system size against SevenNet-Omni~\cite{kim2026sevennetomni}, a public uMLIP trained on \rtwoscan{} data.
The three test systems were fcc Pt, rocksalt LiH (lattice constant \SI{4.083}{\angstrom}), and liquid water (a cubic cell of eight molecules); for each, the number of atoms was increased stepwise by enlarging the supercell.
For each size, the forces were evaluated ten times and the mean time per evaluation was recorded.
Both models were run on a single GPU of the same type (A100 with 80~GB of memory), with the same benchmark script.
SevenNet-Omni was run with the public pretrained model in its matpes\_r2scan task, which corresponds to the \rtwoscan{} functional, using, as in the SevenNet-Omni paper, the implementation with FlashTP~\cite{lee2025flashtp}, which accelerates the tensor-product operations.
The size was increased until the calculation ran out of GPU memory, and the maximum number of atoms was taken as the largest size at which all ten evaluations succeeded.

Table~\ref{tab:speed} summarizes, for the three systems, the maximum number of atoms each model could handle and the time per force evaluation at the largest size SevenNet-Omni could handle.
In all three systems, the maximum number of atoms PFP v9 can handle is five to ten times that of SevenNet-Omni, and the force evaluation at the same size is three to four times faster.
The largest difference is for liquid water: for the 12,288-atom system, the time per evaluation is \SI{0.41}{s} for PFP v9 and \SI{1.68}{s} for SevenNet-Omni, and whereas SevenNet-Omni ran out of GPU memory beyond this size, PFP v9 could handle up to 117,912 atoms.
\begin{table}[htb]
  \centering
  \caption{Maximum number of atoms and inference speed of PFP-R2SCAN and SevenNet-Omni (matpes\_r2scan task), both on a single A100 GPU with 80~GB of memory. The time is that of one force evaluation for the largest system SevenNet-Omni could handle.}
  \label{tab:speed}
  \begin{tabular}{lrrrr}
    \toprule
    & \multicolumn{2}{c}{Maximum atoms} & \multicolumn{2}{c}{Time per force evaluation (s)} \\
    \cmidrule(lr){2-3}\cmidrule(lr){4-5}
    System & PFP-R2SCAN & SevenNet-Omni & PFP-R2SCAN & SevenNet-Omni \\
    \midrule
    Pt fcc & 103,823 & 19,683 & 0.67 & 1.99 \\
    LiH rocksalt & 93,312 & 11,664 & 0.46 & 1.90 \\
    Liquid water & 117,912 & 12,288 & 0.41 & 1.68 \\
    \bottomrule
  \end{tabular}
\end{table}
\section{Conclusion}
\label{sec:conclusion}

PFP v9 extends the training data computed with the \rtwoscan{} functional to the lanthanides from Pr to Lu and the elements from Po to Cm, which PFP v8~\cite{shinagawa2026pfpv8} had deferred, and to the domains, including clusters and adsorption structures, that PFP v8 had lacked, so that PFP-R2SCAN now covers the same 96 elements and the same domains as PFP-PBE and PFP-PBE+U.
The choice of pseudopotentials for the $f$-block elements, left open in PFP v8, was settled by adopting, for the lanthanides from Pr to Yb, the hard potentials that treat all 4$f$ electrons as valence electrons.

In the validation, the formation energies of crystals containing the newly supported elements had an MAE of 93~meV/atom against experimental standard enthalpies of formation, not far from the 81~meV/atom for crystals composed only of previously supported elements.
The adsorption energies of molecules on transition-metal surfaces had an MAE of 354~meV for PFP-R2SCAN and 364~meV for PFP-R2SCAN+D3 against experiment, both at the same level as the 333~meV of DFT-PBE; the chemisorption systems, however, were overbound by 361~meV on average, against 176~meV for DFT-PBE.
Against SevenNet-Omni on the same type of GPU (A100), PFP v9 handled five to ten times as many atoms in all three test systems and evaluated forces three to four times faster.
Taken together, these results show that the extended training data give PFP-R2SCAN the same accuracy for compounds of rare-earth and actinide elements as for the previously supported elements and an overall accuracy for adsorption energies comparable to DFT-PBE, without giving up its advantage in inference speed and system size over SevenNet-Omni.
PFP-R2SCAN can therefore be applied to these systems in the same way as to those covered since PFP v8.

We see three directions for future work.
The first is the selection of Hubbard U parameters for PFP-R2SCAN.
Unlike the PBE-based modes, PFP-R2SCAN has no counterpart with a Hubbard U correction, and extending its applicability to strongly correlated transition-metal oxides and similar systems will require U parameters suited to the \rtwoscan{} functional and a dataset built with them.
The second is surfaces, interfaces, and defects.
Universal MLIPs trained mostly on near-equilibrium bulk structures tend to underestimate surface energies~\cite{deng2025softening,focassio2025surfaces}, as well as defect energies and ion migration barriers~\cite{deng2025softening}.
Although PFP v9 adds adsorption structures to the \rtwoscan{} data, the validation in this technical note covers only adsorption energies on clean low-index surfaces.
Grain boundaries, heterointerfaces, stepped or reconstructed surfaces, and point defects remain to be validated, and the \rtwoscan{} data should be extended to such configurations where the accuracy falls short.
The third is long-range interactions.
Like other universal MLIPs, PFP builds the energy from local atomic environments within a finite cutoff and does not take the total charge or the external field as input, so charged defects, electrified interfaces, electrolytes, and the polarization response of dielectrics lie outside what the present architecture can describe~\cite{anstine2023longrange}.
Treating these systems will require an explicit long-range term, such as learned atomic charges with an electrostatic sum, in addition to the data extensions described above.

\section*{Acknowledgments}

PFP v9 was developed on our in-house supercomputers and on the AI Bridging Cloud Infrastructure (ABCI) 2.0 and 3.0, which are provided by the National Institute of Advanced Industrial Science and Technology (AIST) and AIST Solutions Co., Ltd.
Our use of ABCI 3.0 was supported by the ABCI 3.0 Development Acceleration Program.

\bibliographystyle{unsrtnat}
\bibliography{main}

\end{document}